\documentclass[doublecol]{epl2}
\usepackage{amsfonts,amsmath,amssymb}
\title{Thermalization and Bose--Einstein condensation of quantum light in bulk nonlinear media}
\shorttitle{Thermalization and Bose--Einstein condensation of quantum light in bulk nonlinear media}

\author{
A. Chiocchetta\inst{1}\footnote{These authors contributed equally to this work.} \and
P.-\'E. Larr\'e\inst{2}\footnotemark[\value{footnote}] \and
I. Carusotto\inst{2}
}
\shortauthor{A. Chiocchetta \etal}

\institute{
\inst{1} SISSA --- International School for Advanced Studies and INFN, Via Bonomea 265, 34136 Trieste, Italy \\
\inst{2} INO-CNR BEC Center and Dipartimento di Fisica, Universit\`a di Trento, Via Sommarive 14, 38123 Povo, Italy
}


\abstract{
We study the thermalization and the Bose--Einstein condensation of a paraxial, spectrally narrow beam of quantum light propagating in a lossless bulk Kerr medium. The spatiotemporal evolution of the quantum optical field is ruled by a Heisenberg equation analogous to the quantum nonlinear Schr\"odinger equation of dilute atomic Bose gases. Correspondingly, in the weak-nonlinearity regime, the phase-space density evolves according to the Boltzmann equation. Expressions for the thermalization time and for the temperature and the chemical potential of the eventual Bose--Einstein distribution are found. After discussing experimental issues, we introduce an optical setup allowing the evaporative cooling of a guided beam of light towards Bose--Einstein condensation. This might serve as a novel source of coherent light.
}

\begin{document}

\maketitle

\section{Introduction}

In the last few years, many-body physics has embraced a novel class of systems, the so-called quantum fluids of light \cite{Carusotto2013}. In these optical systems, light and matter combine to generate new photonlike particles that, differently from vacuum photons, are characterized by sizeable effective masses and mutual interactions and, therefore, may give rise to novel states of matter.

One of the most used platforms to study the physics of quantum fluids of light is the semiconductor planar microcavity, in which the cavity photons and the quantum-well excitons strongly couple to form mixed light-matter interacting bosonic quasiparticles called exciton polaritons \cite{Deng2010}. Numerous quantum-hydrodynamics collective phenomena have been investigated theoretically and successfully observed experimentally in such exciton-polariton fluids \cite{Carusotto2013}. Nevertheless, fluids of light in cavity-based systems are inevitably subject to losses, which is typically detrimental for the experimental observation of coherent quantum dynamical features. A more promising configuration for the study of quantum phenomena in fluids of light consists in the paraxial propagation of a quasimonochromatic beam of light in a nonabsorbing bulk nonlinear medium of Kerr type.

It is well known \cite{Boyd1992, Agrawal1995, Rosanov2002} that in such a cavityless, propagating, geometry the complex amplitude of the classical optical field is a slowly varying function of space and time which satisfies a nonlinear wave equation formally identical to the Gross--Pitaevskii (GP) equation of dilute Bose--Einstein (BE) condensates \cite{Pitaevskii2016} after exchanging the roles of the propagation coordinate and of the time parameter. This classical paraxial bulk dynamics may be regarded as the emerging mean-field description of an underlying quantum nonlinear Schr\"odinger dynamics, as formalized in full generality in a recent work by two of us \cite{Larre2015}.

In a recent experimental study \cite{Sun2012}, \textsc{C.~Sun} {\etal} have provided the first observation of classical-wave condensation using a beam of classical monochromatic light propagating in a nonlinear photorefractive crystal. The mechanism underlying this condensation of classical light finds its origin in the thermalization of the classical optical field \cite{Picozzi2007, Lagrange2007, Picozzi2008, Picozzi2008Bis, Barviau2008, Barviau2009, Suret2010, Klaers2010, Klaers2010Bis, Aschieri2011, Michel2011} towards an equilibrium state whose statistics obeys the Rayleigh--Jeans (RJ) thermal law, which corresponds to the classical (high-temperature and/or long-wavelength) limit of the BE distribution.

In this letter, we push this research line forward by investigating the very quantum aspects of the thermalization dynamics of the propagating fluid of light. Making use of the fully quantum theory developed in ref.~\cite{Larre2015}, we discuss the possibility of measuring the Boltzmann tails of the eventual BE distribution, which constitutes the hallmark of the particlelike, quantum, nature of the paraxial beam of light at thermal equilibrium. Inspired by recent advances towards atom-laser devices based on in-waveguide evaporative-cooling schemes \cite{Mandonnet2000, Castin2000, Lahaye2004, Lahaye2005}, we finally propose a mechanism leading to a complete BE condensation in the quantum fluid of light. If realized, such a process would offer a novel route to generate spontaneous optical coherence in a novel concept of coherent-light source.

\section{Quantum formalism}

We consider the propagation in the positive-$z$ direction of a paraxial, spectrally narrow beam of light of central angular frequency $\omega$ in a bulk, electrically neutral, nonmagnetic, nonabsorbing, nonlinear medium of real-valued intensity-dependent refractive index $n_{0}+n_{1}(\mathbf{r}_{\perp},z)+n_{2}\,|\mathcal{E}|^{2}$. Here, $n_{0}$ is the background refractive index, $n_{1}[\mathbf{r}_{\perp}=(x,y),z]$ describes the spatial profile of the refractive index, $n_{2}$ quantifies the strength of the ---spatially local and instantaneous--- Kerr nonlinearity of the medium and $\mathcal{E}$ is the slowly varying \cite{Boyd1992, Agrawal1995, Rosanov2002} envelope of the light wave's electric field $\mathrm{Re}[\mathcal{E}\,\mathrm{e}^{\mathrm{i}(\beta_{0}z-\omega t)}]$ of propagation constant $\beta_{0}=n_{0}\,\omega/c$ in the increasing-$z$ direction, where $c$ denotes the vacuum speed of light. For simplicity's sake, we neglect light polarization and we assume that Raman and Brillouin light-scattering processes on phonons in the optical medium occur at a negligible rate.

Following ref.~\cite{Larre2015}, it is possible to map the quantum propagation of the beam of light in the positive-$z$ direction onto a quantum nonlinear Schr\"odinger evolution of a closed system of many interacting photons in a three-dimensional space spanned by the two-dimensional transverse position vector $\mathbf{r}_{\perp}$ and by the physical time parameter $t$. Introducing the time parameter $\tau=\beta_{1}\,z$ and the three-dimensional position vector $\mathbf{r}=(\mathbf{r}_{\perp},\zeta=t/\beta_{1}-z)$, where $\beta_{1}=\mathrm{d}\beta_{0}/\mathrm{d}\omega=(n_{0}+\omega\,\mathrm{d}n_{0}/\mathrm{d}\omega)/c$ denotes the inverse of the group velocity of the photons in the medium at $\omega$, the quantum mechanical propagation equation of the light beam may be reformulated in the Heisenberg form $\mathrm{i}\,\hbar\,\partial\hat{\Psi}/\partial\tau=[\hat{\Psi},\hat{H}]$, where the quantum field operator $\hat{\Psi}=[c\,\varepsilon_{0}\,n_{0}\,\beta_{1}/(2\,\hbar\,\omega)]^{1/2}\,\hat{\mathcal{E}}$ is the second-quantized slowly varying envelope of the electric field, normalized ($\varepsilon_{0}$ is the vacuum permittivity) in a way to satisfy the usual equal-$\tau$ Bose commutation relations $[\hat{\Psi}(\mathbf{r}_{1},\tau),\hat{\Psi}^{\dag}(\mathbf{r}_{2},\tau)]=\delta^{(3)}(\mathbf{r}_{1}-\mathbf{r}_{2})$ and $[\hat{\Psi}(\mathbf{r}_{1},\tau),\hat{\Psi}(\mathbf{r}_{2},\tau)]=0$, and where
\begin{align}
\notag
\hat{H}&\left.=\int\mathrm{d}^{3}r\;\bigg[\frac{\hbar^{2}}{2\,m_{\perp}}\,\frac{\partial\hat{\Psi}^{\dag}}{\partial\mathbf{r}_{\perp}}\cdot\frac{\partial\hat{\Psi}}{\partial\mathbf{r}_{\perp}}+\frac{\hbar^{2}}{2\,m_{\zeta}}\,\frac{\partial\hat{\Psi}^{\dag}}{\partial\zeta}\,\frac{\partial\hat{\Psi}}{\partial\zeta}\right. \\
\label{Eq:Hamiltonian}
&\left.\hphantom{=}+U(\mathbf{r}_{\perp},\tau)\,\hat{\Psi}^{\dag}\,\hat{\Psi}+\frac{g}{2}\,\hat{\Psi}^{\dag}\,\hat{\Psi}^{\dag}\,\hat{\Psi}\,\hat{\Psi}\bigg]\right.
\end{align}
is the many-body Hamiltonian operator of the system.

In eq.~\eqref{Eq:Hamiltonian}, $U(\mathbf{r}_{\perp},\tau)=-\hbar\,\omega/(c\,\beta_{1})\,n_{1}(\mathbf{r}_{\perp},z)$ is the external potential experienced by the photons, due to the spatial variation of the refractive index, and $g=-2\,(\hbar\,\omega)^{2}/(c^{2}\,\varepsilon_{0}^{\hphantom{2}}\,n_{0}^{\hphantom{2}}\,\beta_{1}^{2})\,n_{2}^{\hphantom{2}}$ is the strength of the effective photon-photon interactions induced by the Kerr nonlinearity.

Even more importantly, $m_{\perp}=\hbar\,\beta_{0}\,\beta_{1}$ and $m_{\zeta}=-\hbar\,\beta_{1}^{3}/\beta_{2}^{\vphantom{3}}$ are the effective masses of the paraxial photons in, respectively, the transverse $\mathbf{r}_{\perp}$ plane and the $\zeta$ direction. In generic media, the values of $m_{\perp,\zeta}$ are typically very different, as they have completely different physical origins: the former originates from paraxial diffraction in the transverse plane while the latter, inversely proportional to the group-velocity-dispersion parameter $\beta_{2}=\mathrm{d}\beta_{1}/\mathrm{d}\omega=(2\,\mathrm{d}n_{0}/\mathrm{d}\omega+\omega\,\mathrm{d}^{2}n_{0}/\mathrm{d}\omega^{2})/c$ of the medium at $\omega$, starts playing a crucial role for nonmonochromatic optical fields having a nontrivial time dependence. Unless the carrier frequency $\omega$ lies in the neighborhood of some optical resonance where dispersion is strong, $m_{\perp}$ is generally much smaller than $m_{\zeta}$; as an example, using tabulated data for fused silica \cite{Malitson1965} around $1.55~\mu\mathrm{m}$ ($1~\mu\mathrm{m}$), one obtains a ratio $m_{\perp}/m_{\zeta}\simeq7\times10^{-3}$ ($m_{\perp}/m_{\zeta}\simeq-8\times10^{-3}$).

As the Hamiltonian \eqref{Eq:Hamiltonian} is only valid within a limited angular-frequency and wavevector range around $(\omega,\beta_{0})$, one has to ensure that photon-photon scattering induces no sizeable photon population outside this paraxial region. Thanks to the conservation of the energy \eqref{Eq:Hamiltonian}, a necessary and ---unless the chromatic dispersion has an unusually complex shape--- sufficient condition is that the two masses $m_{\perp,\zeta}$ have the same sign. The robustness of a coherent photon wave against modulational instabilities imposes further conditions that the longitudinal mass be positive, $m_{\zeta}>0$, and the photon-photon interactions be repulsive, $g>0$; by definition, this amounts to assume that the dielectric is characterized by an anomalous group-velocity dispersion, $\beta_{2}<0$, and a self-defocusing Kerr nonlinearity, $n_{2}<0$ \cite{Larre2015}.

\section{Thermalization time}

In this section, we provide an analytical estimate of the time $\tau_{\mathrm{th}}$ ---that is, of the propagation distance $z_{\mathrm{th}}=\tau_{\mathrm{th}}/\beta_{1}$ along the Kerr medium--- that is necessary for the isolated quantum fluid of light described by the Hamiltonian \eqref{Eq:Hamiltonian} to thermalize. It is worth stressing that the thermalization process is here assumed to occur \textit{via} photon-photon collisions within the fluid only, and not to involve any thermal equilibration with the underlying optical medium, \textit{e.g.}, by photon-phonon scattering or repeated absorption-emission cycles as it was instead the case in the experiment of refs.~\cite{Klaers2010, Klaers2010Bis}.

Assuming for the sake of simplicity that the dielectric is spatially homogeneous, $n_{1}(\mathbf{r}_{\perp},z)=0$, \textit{i.e.}, $U(\mathbf{r}_{\perp},\tau)=0$ in eq.~\eqref{Eq:Hamiltonian}, and that the total interaction energy is small with respect to the total kinetic one in the eventual thermal-equilibrium state, the latter has to be characterized by an occupation number in the plane-wave state of wavevector $\mathbf{k}=[\mathbf{k}_{\perp}=(k_{x},k_{y}),k_{\zeta}]$ and energy $E_{\mathbf{k}}=\hbar_{\vphantom{\perp}}^{2}\,\mathbf{k}_{\perp}^{2}/(2\,m_{\perp}^{\vphantom{2}})+\hbar_{\vphantom{\zeta}}^{2}\,k_{\zeta}^{2}/(2\,m_{\zeta}^{\vphantom{2}})$ of the BE form
\begin{equation}
\label{Eq:BoseEinsteinDistribution}
N_{\mathrm{BE}}(E_{\mathbf{k}},T,\mu)=\bigg[\!\exp\!\bigg(\frac{E_{\mathbf{k}}-\mu}{k_{\mathrm{B}}\,T}\bigg)-1\bigg]^{-1}
\end{equation}
($k_{\mathrm{B}}$ is the Boltzmann constant), where $T$ and $\mu$ are respectively the temperature and the chemical potential of the thermalized quantum fluid of light. As we have assumed there is no thermal contact with the underlying optical medium, $T$ is not related to the temperature of the latter as in refs.~\cite{Klaers2010, Klaers2010Bis}, and both $T$ and $\mu$ are fully determined as functions of the energy and number densities of the photon fluid entering the medium, as detailed in the next section.

A simple model ---based on the quantum nonlinear Schr\"odinger formalism \eqref{Eq:Hamiltonian}--- to investigate the relaxation dynamics of the initial state of the photon fluid, at $\tau=0$ (\textit{i.e.}, $z=0$), towards thermal equilibrium, at $\tau\gtrsim\tau_{\mathrm{th}}$ (\textit{i.e.}, $z\gtrsim z_{\mathrm{th}}$), is provided by the homogeneous [as $U(\mathbf{\mathbf{r}_{\perp}},\tau)=0$] Boltzmann kinetic equation \cite{Griffin2009}
\begin{align}
\notag
\frac{\partial N_{\mathbf{k}}}{\partial\tau}&\left.=\frac{2\,g^{2}}{\hbar}\int\frac{\mathrm{d}^{3}k_{2}}{(2\pi)^{3}}\,\frac{\mathrm{d}^{3}k_{3}}{(2\pi)^{3}}\,\frac{\mathrm{d}^{3}k_{4}}{(2\pi)^{3}}\right. \\
\notag
&\left.\hphantom{=}\times(2\pi)^{3}\,\delta^{(3)}(\mathbf{k}+\mathbf{k}_{2}-\mathbf{k}_{3}-\mathbf{k}_{4})\right. \\
\notag
&\left.\hphantom{=}\times\hphantom{(}2\pi\hphantom{)^{3}}\,\delta(E_{\mathbf{k}}+E_{\mathbf{k}_{2}}-E_{\mathbf{k}_{3}}-E_{\mathbf{k}_{4}})\right. \\
\notag
&\left.\hphantom{=}\times[(N_{\mathbf{k}}+1)\,(N_{\mathbf{k}_{2}}+1)\,N_{\mathbf{k}_{3}}\,N_{\mathbf{k}_{4}}\right. \\
\label{Eq:BoltzmannEquation}
&\left.\hphantom{=}-N_{\mathbf{k}}\,N_{\mathbf{k}_{2}}\,(N_{\mathbf{k}_{3}}+1)\,(N_{\mathbf{k}_{4}}+1)]\right.
\end{align}
for the uniform phase-space density $N_{\mathbf{k}}=N_{\mathbf{k}}(\tau)$ of the paraxial photons occupying the plane-wave state of wavevector $\mathbf{k}$ and energy $E_{\mathbf{k}}$ at the propagation time $\tau$. At long times, \textit{i.e.}, when $\tau\gtrsim\tau_{\mathrm{th}}$, the solution $N_{\mathbf{k}}$ of eq.~\eqref{Eq:BoltzmannEquation} approaches the stationary BE distribution \eqref{Eq:BoseEinsteinDistribution}. Equation \eqref{Eq:BoltzmannEquation} is valid (i) in the absence of condensate and (ii) in the weak-interaction regime. The constraint (i) is satisfied as long as one considers energies and densities yielding noncondensed equilibrium states; otherwise, one has to include the coherent dynamics of the condensate's order parameter in eq.~\eqref{Eq:BoltzmannEquation} \cite{Griffin2009}. The condition (ii) may be checked \textit{a posteriori} by requiring that, in the eventual thermal state, the total interaction energy is small compared to the total kinetic energy, as already supposed in the second paragraph of the present section.

To estimate the effective relaxation time $\tau_{\mathrm{th}}$ towards thermal equilibrium, we are going to mutuate well-known results from the theory of weakly interacting atomic Bose gases. A numerical study \cite{Wu1996} demonstrated that the thermalization time $\tau_{\mathrm{th}}$ of weakly interacting bosonic atoms not too far from thermal equilibrium is typically of the order of $3/\gamma$, where $\gamma$ denotes the average collision rate. This means that about three collisions per particle are sufficient to make the system thermalize. This collision rate may be expressed as \cite{Pitaevskii2016} $\gamma=\rho'\,v'\,\sigma'$, where $\rho'$ denotes the mean number density of the gas, $v'$ is the average norm of the velocity of ideal classical bosons at temperature $T$ and $\sigma'$ is the low-energy boson-boson-scattering cross section. This expression for $\gamma$ holds in the case of an isotropic three-dimensional system. In the present optical case, as highlighted in the previous section, the system is characterized by an anisotropic mass tensor. As a result, the above-given formula for $\gamma$ cannot be applied directly to estimate the time for the quantum fluid of light to relax towards thermal equilibrium.

In order to be able to safely use it, one has to make the kinetic contribution to the Hamiltonian \eqref{Eq:Hamiltonian} isotropic with a common mass $m$ in all the $x$, $y$, $\zeta$ directions. To do so, we introduce the mass parameter $m=(m_{\perp}^{2}\,m_{\zeta}^{\vphantom{2}})^{1/3}$ ---that corresponds to the geometric mean of the paraxial-photon effective masses in the transverse $x$, $y$ and longitudinal $\zeta$ directions--- and the rescaled position vector $\mathbf{r}'=[\mathbf{r}_{\perp}'=(m_{\perp}^{\vphantom{\prime}}/m)^{1/2}\,\mathbf{r}_{\perp}^{\vphantom{\prime}},\zeta'=(m_{\zeta}^{\vphantom{\prime}}/m)^{1/2}\,\zeta]$. As an inversed rescaling holds in momentum space, one readily verifies that such a transformation preserves the spatial as well as the phase-space densities. In the isotropic $\mathbf{r}'$ space, we are then allowed to use the estimate $\tau_{\mathrm{th}}\sim3/\gamma=3/(\rho'\,v'\,\sigma')$ for the thermalization time in terms of the mean number density $\rho'=\rho$, the Boltzmann-averaged velocity $v'=\overline{\hbar\,|\mathbf{k}'|/m}=[8\,k_{\mathrm{B}}\,T/(\pi\,m)]^{1/2}$ and the scattering cross section $\sigma'=8\pi\,a^{\prime2}$, where $a'=m\,g'/(4\pi\,\hbar^{2})$ is the $s$-wave scattering length written as a function of the two-body interaction parameter $g'$ in the isotropic $\mathbf{r}'$ space \cite{Pitaevskii2016}. As our coordinate change preserves both the spatial and the phase-space densities, it is immediate to check that $g'$ is equal to the original photon-photon coupling constant $g$ in the anisotropic $\mathbf{r}$ space, $g'=g$.

Combining the results of the previous paragraph, one eventually gets an explicit formula for the thermalization time $\tau_{\mathrm{th}}$,
\begin{equation}
\label{Eq:ThermalizationTime}
\tau_{\mathrm{th}}\sim3\,\bigg\{\rho\,\bigg[\frac{8\,k_{\mathrm{B}}\,T}{\pi\,(m_{\perp}^{2}\,m_{\zeta}^{\vphantom{2}})^{1/3}}\bigg]^{1/2}\,8\pi\,\bigg[\frac{(m_{\perp}^{2}\,m_{\zeta}^{\vphantom{2}})^{1/3}\,g}{4\pi\,\hbar^{2}}\bigg]^{2}\bigg\}^{-1},
\end{equation}
that corresponds to the usual expression of the thermalization time of a three-dimensional weakly interacting atomic Bose gas, with a mass $(m_{\perp}^{2}\,m_{\zeta}^{\vphantom{2}})^{1/3}$ given by the geometric average of the masses in the transverse $x$, $y$ and longitudinal $\zeta$ directions.

\section{Temperature and chemical potential at thermal equilibrium}

As the kinetic-energy density $E_{\mathrm{kin}}=\int\mathrm{d}^{3}k/(2\pi)^{3}\,N_{\mathbf{k}}\,E_{\mathbf{k}}$ and the photon number density $\rho=\int\mathrm{d}^{3}k/(2\pi)^{3}\,N_{\mathbf{k}}$ are quantities conserved during the evolution of the quantum fluid of light described by eq.~\eqref{Eq:BoltzmannEquation}, the temperature $T$ and the chemical potential $\mu$ characterizing the thermal-equilibrium, at $\tau\gtrsim\tau_{\mathrm{th}}$, BE distribution \eqref{Eq:BoseEinsteinDistribution} may be fixed by the initial, at $\tau=0$, values of $E_{\mathrm{kin}}$ and $\rho$: $E_{\mathrm{kin}}(\tau\gtrsim\tau_{\mathrm{th}})=E_{\mathrm{kin}}(\tau=0)$ and $\rho(\tau\gtrsim\tau_{\mathrm{th}})=\rho(\tau=0)$, where the left-hand sides depend on $T$ and $\mu$ and the right-hand sides are functions of the parameters of the incoming electromagnetic field.

In the final equilibrium state ($\tau\gtrsim\tau_{\mathrm{th}}$, \textit{i.e.}, $z\gtrsim z_{\mathrm{th}}$), using eq.~\eqref{Eq:BoseEinsteinDistribution}, one readily gets $E_{\mathrm{kin}}=\frac{3}{2}\,k_{\mathrm{B}}\,T\,g_{5/2}(f)/(\lambda_{\perp}^{2}\,\lambda_{\zeta}^{\vphantom{2}})$ and $\rho=g_{3/2}(f)/(\lambda_{\perp}^{2}\,\lambda_{\zeta}^{\vphantom{2}})$, where $f=\exp[\mu/(k_{\mathrm{B}}\,T)]$ is the fugacity and $g_{\nu}(f)=\Gamma^{-1}(\nu)\int_{0}^{\infty}\mathrm{d}u~u^{\nu-1}/(f^{-1}\,\mathrm{e}^{u}-1)$ refers to the Bose integral, with $\Gamma(\nu)$ the Euler gamma function. These equations are similar to the well-known results of the ideal Bose gas, with the difference that, in the present optical case, there are two different thermal de Broglie wavelengths $\lambda_{\perp,\zeta}=[2\pi\,\hbar^{2}/(m_{\perp,\zeta}\,k_{\mathrm{B}}\,T)]^{1/2}$ due to the anisotropy of the kinetic energy in eq.~\eqref{Eq:Hamiltonian}. By means of the equation for $E_{\mathrm{kin}}$, one finds that $E_{\mathrm{int}}/E_{\mathrm{kin}}=\frac{1}{3}\,g\,\rho/(k_{\mathrm{B}}\,T)\,g_{3/2}(f)/g_{5/2}(f)$, where $E_{\mathrm{int}}=g\,\rho^{2}/2$ is the mean-field interaction-energy density \cite{Pitaevskii2016} of the fluid of light at equilibrium. Thus, as the Bose integrals $g_{3/2}(f)$ and $g_{5/2}(f)$ are of the same order [$1\leqslant g_{3/2}(f)/g_{5/2}(f)\leqslant 1.94(7)$ for $0\leqslant f\leqslant1$], the weak-interaction condition $E_{\mathrm{int}}\ll E_{\mathrm{kin}}$ required for eq.~\eqref{Eq:BoltzmannEquation} to be valid reads $g\,\rho\ll k_{\mathrm{B}}\,T$, which may be reexpressed in terms of the $s$-wave scattering length $a'$ in the isotropic $\mathbf{r}'$ space as $\rho\,a^{\prime3}\ll(\rho\,\lambda_{\perp}^{2}\,\lambda_{\zeta}^{\vphantom{2}})^{-2}$. Note that this constraint directly implies the usual diluteness condition $\rho\,a^{\prime3}\ll1$ when one enters the quantum-degeneracy regime, $\rho\,\lambda_{\perp}^{2}\,\lambda_{\zeta}^{\vphantom{2}}\gg1$.

We assume that the initial fluid of light ($\tau=0$), \textit{i.e.}, the incident beam of light ($z=0$), is characterized by the following Gaussian distribution in real space:
\begin{equation}
\label{Eq:InitialDistribution}
\langle\hat{\Psi}^{\dag}(\mathbf{r},0)\,\hat{\Psi}(0,0)\rangle=\rho_{0}\,\mathrm{e}^{-\mathbf{r}_{\perp}^{2}/(2\,\ell_{\perp}^{2})}\,\mathrm{e}^{-\zeta^{2}/(2\,\ell_{\zeta}^{2})},
\end{equation}
with finite correlation lengths $\ell_{\perp}$ and $\ell_{\zeta}$ in the transverse $\mathbf{r}_{\perp}$ plane and the longitudinal $\zeta$ direction, respectively. In an actual experiment, the input density $\rho_{0}$ of the quantum fluid of light is tuned by varying the intensity $\mathcal{I}=\hbar\,\omega\,\rho_{0}/\beta_{1}$ of the incoming light beam, the transverse correlation length $\ell_{\perp}$ may be tuned by processing the input beam through spatial light modulators \cite{Sun2012} and the longitudinal correlation length $\ell_{\zeta}$ may in principle be varied by modifying the coherence time $\beta_{1}\,\ell_{\zeta}$ of the incident beam. Note that $\ell_{\perp}$ must be larger than the wavelength $2\pi/\beta_{0}$ of the carrier wave to ensure the paraxiality of the beam of light in the medium and $1/(\beta_{1}\,\ell_{\zeta})$ must be smaller than the frequency range within which the quadratic approximation of the dispersion relation of the medium is valid. Fourier transforming eq.~\eqref{Eq:InitialDistribution} yields the expression of the initial occupation number $N_{\mathbf{k}}(\tau=0)=(2\pi)^{3/2}\,\rho_{0}^{\vphantom{2}}\,\ell_{\perp}^{2}\,\ell_{\zeta}^{\vphantom{2}}\,\mathrm{e}^{-\ell_{\perp}^{2}\,\mathbf{k}_{\perp}^{2}/2}\,\mathrm{e}^{-\ell_{\zeta}^{2}\,k_{\zeta}^{2}/2}$ at $\mathbf{k}$. From this, one obtains, at $\tau=0$, $E_{\mathrm{kin}}=[\hbar_{\vphantom{\perp}}^{\vphantom{-}2}\,\ell_{\perp}^{-2}/m_{\perp}^{\vphantom{-2}}+\hbar_{\vphantom{\zeta}}^{\vphantom{-}2}\,\ell_{\zeta}^{-2}/(2\,m_{\zeta}^{\vphantom{-2}})]\,\rho_{0}$ and $\rho=\rho_{0}$.

Making use of the conservation laws $E_{\mathrm{kin}}(\tau\gtrsim\tau_{\mathrm{th}})=E_{\mathrm{kin}}(\tau=0)$ and $\rho(\tau\gtrsim\tau_{\mathrm{th}})=\rho(\tau=0)$, one eventually gets the following 2-by-2 system:
\begin{equation}
\label{Eq:ConservationLaws}
\frac{3}{2}\,k_{\mathrm{B}}\,T\,\frac{g_{5/2}(f)}{g_{3/2}(f)}=\frac{\hbar_{\vphantom{\perp}}^{\vphantom{-}2}\,\ell_{\perp}^{-2}}{m_{\perp}}+\frac{\hbar_{\vphantom{\zeta}}^{\vphantom{-}2}\,\ell_{\zeta}^{-2}}{2\,m_{\zeta}},\quad\frac{g_{3/2}(f)}{\lambda_{\perp}^{2}\,\lambda_{\zeta}^{\vphantom{2}}}=\rho_{0},
\end{equation}
the resolution of which makes it possible to obtain $T$ and $\mu$ in the final thermal-equilibrium state in terms of $\rho_{0}$, $\ell_{\perp}$, $\ell_{\zeta}$, $m_{\perp}$ and $m_{\zeta}$. Introducing the effective temperatures $T_{\perp,\zeta}=2\pi\,\hbar^{2}/(k_{\mathrm{B}}\,m_{\perp,\zeta}^{\vphantom{2}}\,\ell_{\perp,\zeta}^{2})$, the first of eqs.~\eqref{Eq:ConservationLaws} may be rewritten as $6\pi\,g_{5/2}(f)/g_{3/2}(f)\,T=2\,T_{\perp}+T_{\zeta}$, which shows that the transverse and longitudinal modes, initially distributed at different temperatures $T_{\perp}\neq T_{\zeta}$, eventually equilibrate at the same temperature $T$.

\section{Experimental considerations}

Reminding the definition of the spatial coordinate $\zeta$, the third component of the paraxial-photon wavevector $\mathbf{k}$ may be expressed as \cite{Larre2015, Larre2016} $k_{\zeta}=-\beta_{1}\,\Delta\omega$, where $\Delta\omega$ is the detuning from the angular frequency $\omega$ of the pump. As a result, the measurement of the BE distribution \eqref{Eq:BoseEinsteinDistribution} as a function of $\mathbf{k}=(\mathbf{k}_{\perp},-\beta_{1}\,\Delta\omega)$ requires a good angular resolution to isolate the light deflected with a transverse wavevector $\mathbf{k}_{\perp}$ as well as a good spectral resolution to isolate the angular-frequency component of the transmitted light at $\omega\pm\Delta\omega$.

On the other hand, to have access to the large-momentum, Boltzmann, tails of the BE distribution ---and so, in turn, to the whole BE distribution as a function of $\mathbf{k}$--- at the exit face of the nonlinear dielectric where the fluid of light is imaged, some conditions have to be satisfied.

The inverse of the de Broglie wavelengths $\lambda_{\perp}$ and $\lambda_{\zeta}$ being the natural scales of variation of $N_{\mathrm{BE}}(E_{\mathbf{k}},T,\mu)$ as a function of $\mathbf{k}_{\perp}$ and $k_{\zeta}$, a first condition for detecting the whole BE distribution in the transmitted beam of light is that $\lambda_{\perp}$ and $\lambda_{\zeta}$ must verify the constraints satisfied respectively by $\ell_{\perp}$ and $\ell_{\zeta}$ (see the third paragraph of the previous section).

A second, perhaps more challenging, condition concerns the length of the bulk nonlinear medium, which has to be at least of the order of the distance $z_{\mathrm{th}}=\tau_{\mathrm{th}}/\beta_{1}$ necessary for the quantum fluid of light to fully relax towards thermal equilibrium. Making use of the analytical result \eqref{Eq:ThermalizationTime} and of the first of eqs.~\eqref{Eq:ConservationLaws} with the reasonable estimate $g_{5/2}(f)\sim g_{3/2}(f)$ for $0\leqslant f\leqslant1$, one finds that $z_{\mathrm{th}}$ must behave at a given carrier wave at $(\omega,\beta_{0}=n_{0}\,\omega/c)$ as
\begin{equation}
\label{Eq:ThermalizationDistance}
z_{\mathrm{th}}=\frac{K}{|n_{2}|^{2}\,\mathcal{I}}\,\bigg[\frac{|\beta_{2}|}{\ell_{\perp}^{-2}+\beta_{0}^{\vphantom{2}}\,|\beta_{2}^{\vphantom{2}}|/(2\,\beta_{1}^{2})\,\ell_{\zeta}^{-2}}\bigg]^{1/2},
\end{equation}
where $K$ depends on $\hbar$, $c$, $\varepsilon_{0}$, $k_{\mathrm{B}}$ and on $\omega$, $n_{0}$, $\beta_{0}$. As a most important contribution, it is immediate to see that the stronger the Kerr nonlinearity is, the shorter the thermalization distance $z_{\mathrm{th}}$ is. Plugging explicit values into \eqref{Eq:ThermalizationDistance}, we estimate for a light beam of $1.55~\mu\mathrm{m}$ wavelength, $1~\mathrm{W}/\mu\mathrm{m}^{2}$ intensity and initial $\beta_{0}\,\ell_{\perp,\zeta}=10$ coherence lengths propagating in bulk silica [$2\,|n_{2}|/(c\,\varepsilon_{0}\,n_{0})\sim10^{-20}~\mathrm{m}^{2}/\mathrm{W}$] an unreasonably long $z_{\mathrm{th}}\sim10^{13}~\mathrm{m}$, \textit{i.e.}, of the order of the estimated radius of the solar system{\dots}

While an experiment using such standard bulk nonlinear media looks clearly unfeasable, very promising alternatives are offered by resonant media where photons are strongly mixed with matter excitations. In this way, very strong effective photon-photon interactions may be obtained, \textit{e.g.}, for polaritons in bulk semiconducting materials showing narrow exciton lines such as GaAs or ZnSe \cite{Carusotto2013}. This effect can be further reinforced by many orders of magnitude if the chosen material excitation involves spatially wide (even almost micron-sized) Rydberg states, either in optically dressed atomic gases in the so-called Rydberg-EIT regime \cite{Peyronel2012} or in highest-quality solid-state $\mathrm{Cu_{2}O}$ samples \cite{Jang2011}. A further advantage of resonant media is the wide tunability of the optical parameters simply by changing the carrier frequency $\omega$, which is of a great utility to ensure the dynamical stability of the photon fluid.

\section{Discussion of a recent experiment}

In ref.~\cite{Sun2012}, \textsc{C.~Sun} {\etal} reported having experimentally observed the relaxation of a classical, \textit{i.e.}, not quantum, fluid of interacting photons towards a thermal-equilibrium state. A beam of classical monochromatic light, initially prepared in a nonthermal state \textit{via} a suitable tayloring of the incident phase profile, was made to propagate in a photorefractive crystal whose optical nonlinearity was strong enough to make the transverse angular distribution of the beam of light fastly evolve towards a RJ-type, \textit{i.e.}, classical, thermal law. For small enough initial kinetic energies, a marked peak around $\mathbf{k}_{\perp}=0$ was observed in the transverse-momentum-$\mathbf{k}_{\perp}$ distribution, which was interpreted as a signature of the occurrence of a kinetic condensation of classical waves.

In order to fully understand the analogies and the differences with our quantum study, we can start by noting that a key conceptual assumption of the experiment \cite{Sun2012} is that the light beam remains perfectly monochromatic all along its propagation across the nonlinear crystal. Under a mean-field approximation and provided no spontaneous temporal modulations such as self-pulsing \cite{Larre2015} occur, monochromaticity at all distances is a trivial consequence of the classical GP form of the nonlinear Schr\"odinger field equation corresponding to the quantum Hamiltonian \eqref{Eq:Hamiltonian}.

On the other hand, monochromaticity corresponds within the framework of our quantum theory to having at all propagation times $\tau$ a factorized momentum distribution $N_{\mathbf{k}}(\tau)=N_{\mathbf{k}_{\perp}}(\tau)\,N_{k_{\zeta}}$, where the transverse-momentum distribution $N_{\mathbf{k}_{\perp}}(\tau)$ evolves with $\tau$ while the longitudinal one $N_{k_{\zeta}}$ remains constant and proportional to the Dirac function $\delta(k_{\zeta})$ at all $\tau$'s. Monochromaticity at all $\tau$'s then requires that no scattering process can change the $k_{\zeta}$'s of the colliding paraxial photons.

Most remarkably, the specific form of the optical nonlinearity of the photorefractive crystal used in the experiment \cite{Sun2012} automatically serves this purpose, as its slow response involves the time-$t$ average of the optical intensity and ---in many-body terms--- corresponds to infinite-range interactions along the $\zeta$ axis. As a result, all processes that would generate frequencies different from the incident one are suppressed. Keeping in mind that the population is sharply peaked on the only occupied states with $k_{\zeta}=0$, it is then straightforward to see that the kinetics will eventually relax to the classical RJ distribution $N_{\mathrm{RJ}}(E_{\mathbf{k}},T,\mu)=k_{\mathrm{B}}\,T/(E_{\mathbf{k}}-\mu)$ rather than to the BE one \eqref{Eq:BoseEinsteinDistribution}: because of the $\delta$-shaped factor $N_{k_{\zeta}}$ in the $N_{\mathbf{k}}$'s, all the quantum ``$+\,1$'' terms in the Boltzmann equation \eqref{Eq:BoltzmannEquation} are in fact irrelevant, so that the quantum kinetics reduces to a classical one.

The situation is of course completely different if a local and instantaneous nonlinearity is used in an experiment. Within our theory \cite{Larre2015}, this corresponds to a local interaction in the three-dimensional $x$, $y$, $\zeta$ space. As a result, wave-mixing processes can mix all the three components of the momentum, therefore allowing for a full three-dimensional thermalization of the photon gas in both its transverse-momentum-$\mathbf{k}_{\perp}$ distribution and its physical-frequency-$\Delta\omega$ distribution, where $\Delta\omega=-k_{\zeta}/\beta_{1}$ is measured from the carrier wave at $\omega$. Given the quantum nature of our model, the final result of this thermalization process will be a BE distribution of the form \eqref{Eq:BoseEinsteinDistribution}, which automatically solves all the ultraviolet black-body catastrophes that infest classical theories such as the one used in ref.~\cite{Sun2012}. As a final point, it is worth highlighting that thermalization to a quantum distribution is based on the quantum ``$+\,1$'' terms in the Boltzmann equation and thus does not benefit from the large Bose stimulation factor involved in the thermalization of classical waves. Together with the typically weaker Kerr optical nonlinearity of fast media, this explains why our prediction for $z_{\mathrm{th}}$ is dramatically longer than the experimental one of ref.~\cite{Sun2012}.

\section{Evaporative cooling and BE condensation of a beam of light}

An interesting consequence of the above-investigated thermalization process appears when the quantum fluid of light enters the BE-condensed phase. From the theory of the ideal Bose gas \cite{Pitaevskii2016}, the critical line for BE condensation in the $(\rho_{0},T)$ plane may be obtained by imposing $f=1$ in the second of the thermal-state equations \eqref{Eq:ConservationLaws}, which yields the usual formula for the BE-condensation critical temperature,
\begin{equation}
\label{Eq:BECTemperature}
T_{\mathrm{c}}=\frac{2\pi\,\hbar^{2}}{k_{\mathrm{B}}\,(m_{\perp}^{2}\,m_{\zeta}^{\vphantom{2}})^{1/3}}\,\bigg[\frac{\rho_{0}}{\zeta(3/2)}\bigg]^{2/3},
\end{equation}
in terms of the Riemann zeta function at $3/2$, $\zeta(3/2)=g_{3/2}(1)=2.61(2)$, and the before-introduced geometric mean $m=(m_{\perp}^{2}\,m_{\zeta}^{\vphantom{2}})^{1/3}$ of the paraxial-photon effective masses. To realize a BE condensate of light in a bulk geometry, the experimentalist has to choose the rescaled intensity $\rho_{0}$ and the correlation lengths $\ell_{\perp}$ and $\ell_{\zeta}$ of the incident beam in such a way that the temperature $T$ in the thermal state, solution of eqs.~\eqref{Eq:ConservationLaws}, is smaller than $T_{\mathrm{c}}$ given by eq.~\eqref{Eq:BECTemperature}.

Following the theoretical and experimental investigations \cite{Mandonnet2000, Castin2000, Lahaye2004, Lahaye2005} of the evaporative cooling of an atomic beam propagating in a magnetic trap, a promising way to facilitate BE condensation in the quantum fluid of light consists in progressively making the photon beam evaporate in the transverse $\mathbf{r}_{\perp}=(x,y)$ directions.

\begin{figure}[t!]
\centering
\includegraphics[width=\linewidth]{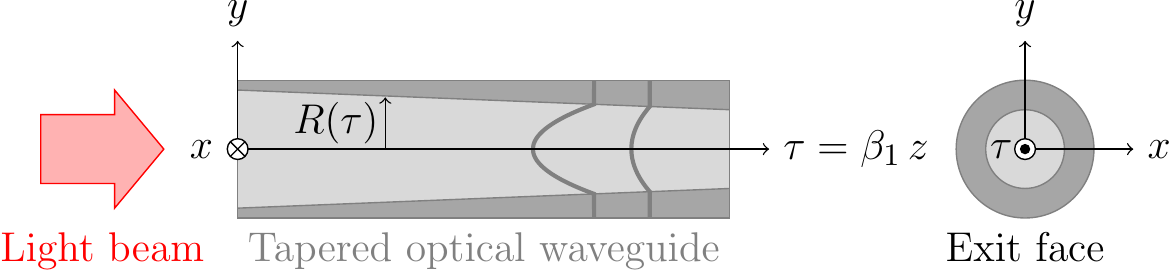}
\caption{(Color online) Sketch of an optical platform allowing the evaporative cooling of a quantum fluid of light (red) to temperatures below the critical temperature for BE condensation. The core (light gray) of radius $R(\tau)$ and the cladding (dark gray) of the waveguide are designed so that the photons are trapped in an effective harmonic potential (thick gray curves) whose maximum amplitude $\propto R^{2}(\tau)$ diminishes as the propagation time $\tau$ increases. This removes the high-energy photons from the fluid of light, which then cools down.}
\label{Fig:EvaporativeCooling}
\end{figure}

This can be obtained by introducing a time-dependent trapping potential $U(\mathbf{r}_{\perp},\tau)\neq0$ into the Hamiltonian \eqref{Eq:Hamiltonian}, for instance of truncated harmonic form $U(\mathbf{r}_{\perp},\tau)=\frac{1}{2}\,m_{\perp}^{\vphantom{2}}\,\omega_{\perp}^{2}\,\mathbf{r}_{\perp}^{2}$ for $|\mathbf{r}_{\perp}|\leqslant R(\tau)$ and $U(\mathbf{r}_{\perp},\tau)=\frac{1}{2}\,m_{\perp}^{\vphantom{2}}\,\omega_{\perp}^{2}\,R^{2}(\tau)$ for $|\mathbf{r}_{\perp}|>R(\tau)$, where the radius $R(\tau)$ is a decreasing function of the propagation time $\tau$. Based on the relation $n_{1}(\mathbf{r}_{\perp},z)=-c\,\beta_{1}/(\hbar\,\omega)\,U(\mathbf{r}_{\perp},\tau)$ between the spatial profile of the refractive index and the effective potential in eq.~\eqref{Eq:Hamiltonian}, this truncated harmonic trap  may be realized by means of a conically tapered multimode optical waveguide, as pictorially sketched in fig.~\ref{Fig:EvaporativeCooling}. The core [$|\mathbf{r}_{\perp}|\leqslant R(\beta_{1}\,z)$] is taken to have an inverse-parabolic refractive-index profile, while the cladding [$|\mathbf{r}_{\perp}|>R(\beta_{1}\,z)$] is homogeneous with a refractive index smoothly connecting the one of the core's edge.

As a result of this tapering, the maximum value of the trapping potential decreases as $\tau$ increases, so that the large-momentum (or large-energy) tails of the photon distribution are progressively removed. At the same time, the remaining photons keep reequilibrating to lower and lower temperatures under the effect of collisions, until the fluid of light eventually crosses the critical temperature for BE condensation.

Upon the $t\longleftrightarrow z$ mapping, BE condensation from an initially thermal photon gas corresponds to the appearance of spontaneous optical coherence when an initially incoherent beam of light is injected into the nonlinear medium: the long-range order of the BE condensate of light reflects into optical coherence extending for macroscopically long times $t$ and distances $x$, $y$. In contrast to trivial angular- and frequency-filtering processes, a key element of our proposal are the collisions between the photons, that allow the fluid of light to reestablish thermal equilibrium at lower and lower temperatures while the most energetic photons keep being removed.

\section{Conclusion}

In this letter, we have investigated the relaxation dynamics of a paraxial, quasimonochromatic beam of quantum light towards thermal equilibrium in a lossless bulk Kerr medium. Following ref.~\cite{Larre2015}, the propagation of the quantum light field has been mapped onto a quantum nonlinear Schr\"odinger evolution of a conservative quantum fluid of many interacting bosons. Correspondingly, in the weak-interaction regime, the evolution of the momentum distribution from an arbitrary nonthermal state towards a thermal state with a BE form can be modeled by the Boltzmann kinetic equation, which offers analytical formulas for the thermalization time and for the final temperature and chemical potential in terms of the parameters of the input beam and of the medium.

In addition to extending the concept of classical-light-wave condensation \cite{Sun2012} to a fully quantum level and solving well-known ultraviolet pathologies of existing classical theories, our results suggest an intriguing long-term application as a novel source of coherent light: taking inspiration from related advances in atom-laser devices \cite{Mandonnet2000, Castin2000, Lahaye2004, Lahaye2005}, we have pointed out a novel in-waveguide evaporative-cooling scheme to obtain spontaneous macroscopic optical coherence from an initially incoherent beam of light. As our proposal does not rely on population-inverted atomic transitions, it holds the promise of being implemented in an arbitrary domain of the electromagnetic spectrum.

\acknowledgments

We are grateful to \textsc{G.~Ferrari}, \textsc{A.~Gambassi}, \textsc{A.~Minguzzi}, \textsc{C.~Miniatura} and \textsc{M.~Richard} for helpful and stimulating discussions. \textsc{AC} acknowledges the kind hospitality of the BEC Center, where this work was partially done. \textsc{P-\'EL} and \textsc{IC} were funded by the EU-FET Proactive Grant AQuS, Project No.~640800, and by the Provincia Autonoma di Trento, partly through the Call ``Grandi Progetti 2012'', Project SiQuro.


\begin{thebibliography}{99}

\bibitem{Carusotto2013}
\Name{Carusotto I. \and Ciuti C.}
\Review{Rev. Mod. Phys.}
\Vol{85}
\Year{2013}
\Page{299}

\bibitem{Deng2010}
\Name{Deng H. \etal}
\Review{Rev. Mod. Phys.}
\Vol{82}
\Year{2010}
\Page{1489}

\bibitem{Boyd1992}
\Name{Boyd R. P.}
\Book{Nonlinear Optics}
\Publ{Academic Press, San Diego}
\Year{1992}

\bibitem{Agrawal1995}
\Name{Agrawal G. P.}
\Book{Nonlinear Fiber Optics}
\Publ{Academic Press, San Diego}
\Year{1995}

\bibitem{Rosanov2002}
\Name{Rosanov N. N.}
\Book{Spatial Hysteresis and Optical Patterns}
\Publ{Springer, Berlin}
\Year{2002}

\bibitem{Pitaevskii2016}
\Name{Pitaevskii L. P. \and Stringari S.}
\Book{Bose--Einstein Condensation and Superfluidity}
\Publ{Oxford Science Publications, Oxford}
\Year{2016}

\bibitem{Larre2015}
\Name{Larr\'e P.-\'E. \and Carusotto I.}
\Review{Phys. Rev. A}
\Vol{92}
\Year{2015}
\Page{043802}

\bibitem{Sun2012}
\Name{Sun C. \etal}
\Review{Nat. Phys.}
\Vol{8}
\Year{2012}
\Page{470}

\bibitem{Picozzi2007}
\Name{Picozzi A.}
\Review{Opt. Express}
\Vol{15}
\Year{2007}
\Page{9063}

\bibitem{Lagrange2007}
\Name{Lagrange S. \etal}
\Review{EPL}
\Vol{79}
\Year{2007}
\Page{64001}

\bibitem{Picozzi2008}
\Name{Picozzi A.}
\Review{Opt. Express}
\Vol{16}
\Year{2008}
\Page{17171}

\bibitem{Picozzi2008Bis}
\Name{Picozzi A. \and Rica S.}
\Review{EPL}
\Vol{84}
\Year{2008}
\Page{34004}

\bibitem{Barviau2008}
\Name{Barviau B. \etal}
\Review{Opt. Lett.}
\Vol{33}
\Year{2008}
\Page{2833}

\bibitem{Barviau2009}
\Name{Barviau B. \etal}
\Review{Opt. Express}
\Vol{17}
\Year{2009}
\Page{7392}

\bibitem{Suret2010}
\Name{Suret P. \etal}
\Review{Phys. Rev. Lett.}
\Vol{104}
\Year{2010}
\Page{054101}

\bibitem{Klaers2010}
\Name{Klaers J. \etal}
\Review{Nat. Phys.}
\Vol{6}
\Year{2010}
\Page{512}

\bibitem{Klaers2010Bis}
\Name{Klaers J. \etal}
\Review{Nature}
\Vol{468}
\Year{2010}
\Page{545}

\bibitem{Aschieri2011}
\Name{Aschieri P. \etal}
\Review{Phys. Rev. A}
\Vol{83}
\Year{2011}
\Page{033838}

\bibitem{Michel2011}
\Name{Michel C. \etal}
\Review{Phys. Rev. A}
\Vol{84}
\Year{2011}
\Page{033848}

\bibitem{Mandonnet2000}
\Name{Mandonnet E. \etal}
\Review{Eur. Phys. J. D}
\Vol{10}
\Year{2000}
\Page{9}

\bibitem{Castin2000}
\Name{Castin Y. \etal}
\Review{J. Mod. Opt.}
\Vol{47}
\Year{2000}
\Page{2671}

\bibitem{Lahaye2004}
\Name{Lahaye T. \etal}
\Review{Phys. Rev. Lett.}
\Vol{93}
\Year{2004}
\Page{093003}

\bibitem{Lahaye2005}
\Name{Lahaye T. \etal}
\Review{Phys. Rev. A}
\Vol{72}
\Year{2005}
\Page{033411}

\bibitem{Malitson1965}
\Name{Malitson I. H.}
\Review{J. Opt. Soc. Am.}
\Vol{55}
\Year{1965}
\Page{1205}

\bibitem{Griffin2009}
\Name{Griffin A. \etal}
\Book{Bose-Condensed Gases at Finite Temperature}
\Publ{Cambridge University Press, Cambridge}
\Year{2009}

\bibitem{Wu1996}
\Name{Wu H. \and Foot C. J.}
\Review{J. Phys. B: At. Mol. Opt. Phys.}
\Vol{29}
\Year{1996}
\Page{L321}

\bibitem{Larre2016}
\Name{Larr\'e P.-\'E. \and Carusotto I.}
\Review{Eur. Phys. J. D}
\Vol{70}
\Year{2016}

\bibitem{Peyronel2012}
\Name{Peyronel T. \etal}
\Review{Nature}
\Vol{488}
\Year{2012}
\Page{57}

\bibitem{Jang2011}
\Name{Jang J. I.}
\Book{Optoelectronics: Materials and Techniques}
\Publ{InTech}
\Year{2011}%
, Chap.~5

\end{thebibliography}
\end{document}